\documentclass[10pt,letterpaper,twocolumn]{article}
\usepackage[pagenumbers]{cvpr} % To force page numbers, e.g. for an arXiv version

\definecolor{cvprblue}{rgb}{0.21,0.49,0.74}
\usepackage[pagebackref,breaklinks,colorlinks,allcolors=cvprblue]{hyperref}
\usepackage{bm}
\usepackage{amsmath}
\usepackage{graphicx}
\usepackage[linesnumbered, ruled, lined]{algorithm2e}
\usepackage{multirow}
\usepackage{placeins}
\usepackage{utfsym}
\usepackage{cuted}      
\usepackage{caption} 
\usepackage{float}

\usepackage{algorithmic}
\usepackage{amssymb}
\DeclareFixedFont{\ttb}{T1}{txtt}{bx}{n}{8} % for bold
\DeclareFixedFont{\ttm}{T1}{txtt}{m}{n}{8}  % for normal

\usepackage{color}
\definecolor{deepblue}{rgb}{0,0,0.5}
\definecolor{deepred}{rgb}{0.6,0,0}
\definecolor{deepgreen}{rgb}{0,0.5,0}

\def\paperID{6148} % *** Enter the Paper ID here
\def\confName{CVPR}
\def\confYear{2025}

\title{Towards Geometric and Textural Consistency 3D Scene Generation via Single Image-guided Model Generation and Layout Optimization}
\author{
Xiang Tang$^{1,2}$ \quad
Ruotong Li$^{2}$ \quad
Xiaopeng Fan$^{3,2,4}$ \\
$^1$Harbin Institute of Technology, Shenzhen \quad $^2$Peng Cheng Laboratory \\ 
$^3$Harbin Institute of Technology \quad $^4$Harbin Institute of Technology, Suzhou Research Institute\\
}
\begin{document}

\twocolumn[{
\renewcommand\twocolumn[1][]{#1}
\maketitle
}]

\begin{abstract}
In recent years, 3D generation has made great strides in both academia and industry. However, generating 3D scenes from a single RGB image remains a significant challenge, as current approaches often struggle to ensure both object generation quality and scene coherence in multi-object scenarios. To overcome these limitations, we propose a novel three-stage framework for 3D scene generation with explicit geometric representations and high-quality textural details via single image-guided model generation and spatial layout optimization. Our method begins with an image instance segmentation and inpainting phase, which recovers missing details of occluded objects in the input images, thereby achieving complete generation of foreground 3D assets. Subsequently, our approach captures the spatial geometry of reference image by constructing a pseudo-stereo viewpoint for camera parameter estimation and scene depth inference, while employing a model selection strategy to ensure optimal alignment between the 3D assets generated in the previous step and the input. Finally, through model parameterization and minimization of the Chamfer distance between point clouds in 3D and 2D space, our approach optimizes layout parameters to produce an explicit 3D scene representation that maintains precise alignment with the input guidance image. Extensive experiments on multi-object scene image sets have demonstrated that our approach not only outperforms state-of-the-art methods in terms of geometric accuracy and texture fidelity of individual generated 3D models, but also has significant advantages in scene layout synthesis. Project page: \url{https://xdlbw.github.io/sing3d/}

\end{abstract}

\section{INTRODUCTION}

3D generation enables the intuitive and rapid creation of immersive, photorealistic objects and environments, releasing artists and designers from labor-intensive manual work. As a pivotal research direction in computer graphics, it holds significant application value for digital content creation, virtual reality, autonomous navigation, and embodied intelligence. Rapid advancements in 3D content generation have not only enriched 3D representations \cite{mildenhall2021nerf, kerbl20233d, shen2021deep}, but also enhanced training of feedforward generative approaches through the establishment of large-scale datasets \cite{chang2015shapenet, wu20153d, deitke2023objaverse, fu20213d}. In addition, deep generative architectures represented by Generative Adversarial Networks \cite{gao2022get3d, barthel2024gaussian} and diffusion models \cite{poole2022dreamfusion, yu2023text, liang2024luciddreamer, wang2023prolificdreamer, ma2024scaledreamer, tang2025recent} have demonstrated remarkable capabilities in modeling complex visual distributions, enabling highly efficient 3D generation.

When focusing on the specialized domain of single RGB image-to-3D reconstruction, the uncertainty inherent in single-view often leads to severe geometric ambiguities and insufficient reconstruction of occluded regions, resulting in incomplete geometries or inconsistent textures. Although current methods \cite{qian2023magic123, xu2023dmv3d, xiang2024structured} have demonstrated remarkable results in single object generation, their performance significantly deteriorates when handling scenes with complex compositions of multiple objects. These approaches treat objects that are obscured by each other as a single entity while entangling truly separated instances, which leads to issues such as loss of details, incomplete scene composition, and multi-view inconsistency in the generated results. Furthermore, even though there are works on compositional scene synthesis \cite{rahamim2024lay, zhou2024zero}, the absence or erroneous estimation of depth information frequently leads to abnormal object placement and orientation. The reason lies in the fact that monocular inputs complicate the accurate estimation of camera parameters and scene depth, which are crucial for predicting inter-object spatial relationships and optimizing their layout within generated scenes.

Motivated by these observations, our work focuses on improving the geometry representation of 3D generation by extracting the correct object instances from the reference image and ensuring the accuracy of the synthesized scene layout through high-quality scene depth estimation at the same time. To achieve this, our work designs a decomposition-recomposition strategy, which first achieves independent object generation through instance decoupling from the input image, followed by spatial relationship reconstruction via layout optimization. This framework not only leverages the full potential of existing single-object generation models but also effectively resolves generation challenges arising from multi-object interactions and occlusions. 

\begin{figure*}[t]
    \centering
    \vspace{-0.4cm}
    \includegraphics[width=\textwidth]{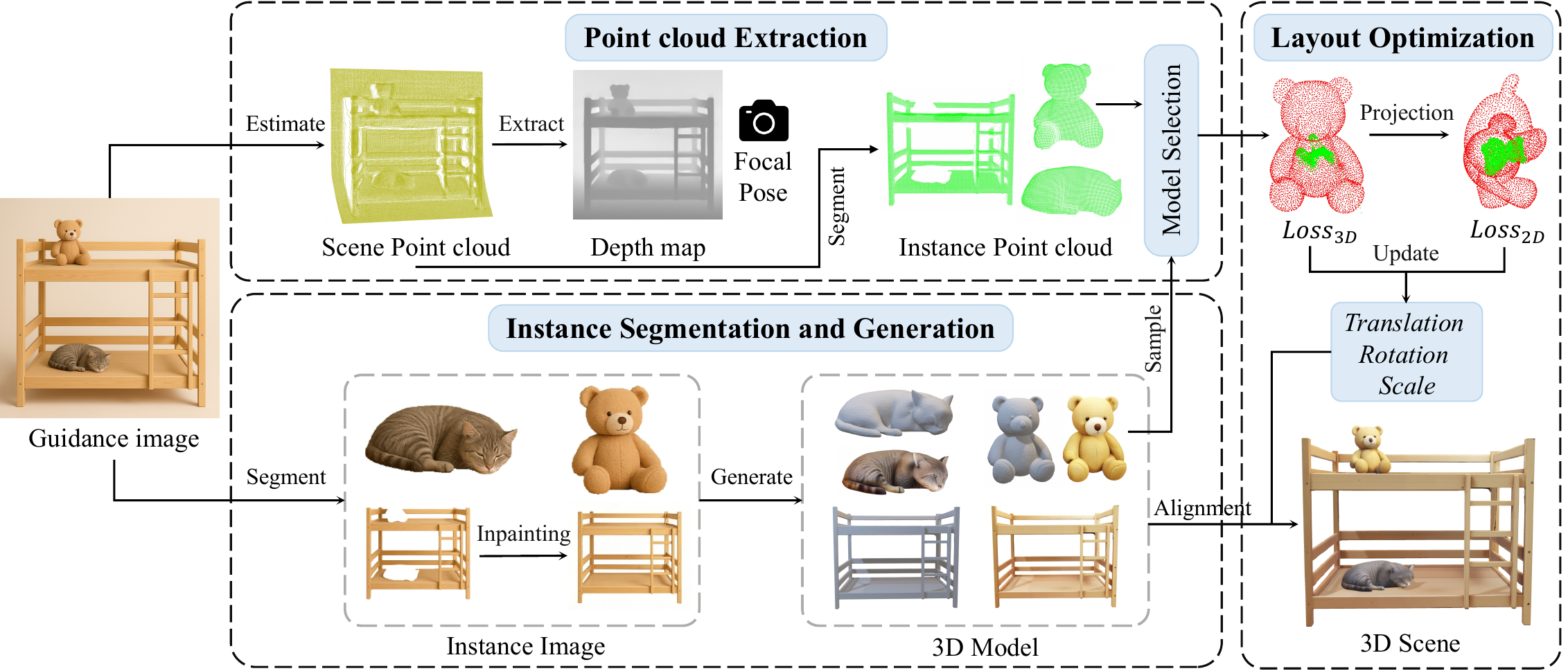}
    \vspace{-0.6cm}
    \caption{\textbf{Overview of our method.} Our approach accomplishes complex scene generation through three collaborative subtasks. Given a single image as guidance, during the Instance Segmentation and Generation stage, we first perform object detection and instance segmentation to obtain instance-specific images, masks, and related information. After that, we focus on repairing imperfect instance images (e.g., bed) and generates corresponding multiple 3D assets with a generative model. In the Point Cloud Extraction stage, we estimate the depth map, camera parameters, and complete scene point cloud of the input image, and perform segmentation with the previously obtained masks to derive an independent point cloud representation for each instance. Additionally, we sample the generated 3D models into point clouds and implement a model selection strategy to choose 3D assets that best match the instance images. During layout optimization, we optimize layout parameters by minimizing the 3D and 2D Chamfer Distance between the optimal model point cloud (depicted in \textcolor{red}{red}) and the instance point cloud (depicted in \textcolor{green}{green}), finally constructing a 3D scene that maintains high consistency with the reference image layout.}
    \vspace{-0.4cm}
    \label{Overview}
\end{figure*}

As shown in Fig. \ref{Overview}, we adopt a divide-and-conquer philosophy, decomposing the generation pipeline into three collaborative subtasks: 1) \textbf{Instance Segmentation and Generation} performs object detection and instance segmentation on the input image to obtain segmented instances, masks, semantic labels, and confidence scores, subsequently refining imperfect instance images and producing multiple high-fidelity 3D models for each object. 2) \textbf{Point Cloud Extraction} estimates camera parameters and scene depth through pseudo-stereo vision, subsequently extracts both global scene and individual instance point clouds, and adopts a model selection strategy to identify the 3D assets that best match the instance images. 3) \textbf{Layout Optimization} parameterizes 3D instances and optimizes their spatial arrangements through point cloud matching, ensuring compositional consistency in the final synthesized scene. We construct a small dataset containing multi-object scenarios' images for method validation. The experimental results show that when processing images with significant object occlusions and intricate spatial relationships, our method not only maintains superior single-object generation quality but also achieves precise scene layout recovery, showing substantial improvements over prior approaches. Our main contributions can be summarized as follows:
\begin{itemize}
\item[$\bullet$]We propose a modular three-stage framework that can extract multiple independent 3D assets with explicit geometry representation and high-quality textural details from a single image, together with accurate scene layout recovery.
\item[$\bullet$] We devise an asset generation-selection strategy that integrates image inpainting and model matching in order to effectively overcome the insufficient object reconstruction caused by occlusions, thus enabling our method to produce 3D assets that best match the objects in reference images.
\item[$\bullet$] We introduce a novel layout optimization technique that leverages object point cloud representations obtained from subtask 2 to compute 3D Chamfer Distance and 2D projection space loss, effectively ensuring geometric and spatial consistency between the generated 3D scene and the original 2D input.
\end{itemize}

\section{RELATED WORK}

\subsection{Image-based 3D Generation}

The controllable generation of 3D objects from images has been significantly enhanced by leveraging strong visual priors. Early research centered on Score Distillation Sampling (SDS) \cite{poole2022dreamfusion}, with representative methods \cite{melas2023realfusion, tang2023make, tang2023dreamgaussian} exploiting knowledge from pretrained 2D diffusion models to optimize 3D representations. However, inherent geometric ambiguities in single-view inputs frequently lead to multi-faced Janus problems. To address this bottleneck, subsequent studies \cite{liu2023zero, shi2023zero123++, liu2023one, liu2024one, liu2023syncdreamer, long2024wonder3d, li2025caddreamer, wu2024unique3d} have focused on fine-tuning diffusion models to generate multi-view consistent images for 3D generation. Recent advancements exemplified by 3DTopia-XL \cite{chen2025primx} and Trellis \cite{xiang2024structured} integrate VAEs with Transformer architectures, demonstrating exceptional geometric and textural fidelity in cross-modal (i.e., text/image-to-3D) generation tasks. 

While these methods demonstrate remarkable performance in single object generation, they often struggle with multi-object scenes. In the domain of indoor multi-instance scene synthesis, works such as \cite{song2023roomdreamer, li2024genrc, schult2024controlroom3d, zhou2024dreamscene360, li2024scenedreamer360} have explored generation paradigms based on panoramic image priors, yet they are constrained by the inability to decouple individual objects. One of our main objectives in this work is to extend the object-level generation to synthesize complex scenes.

\subsection{Compositional 3D Scene Synthesis}

Compositional scene synthesis aims to arrange 3D assets through spatial-topological relationships between objects specified in layouts to generate scenes. Previous approaches primarily rely on spatial relationship priors \cite{wang2018deep, zhao2024roomdesigner} and 3D bounding boxes \cite{po2024compositional} to constrain object placement, but their dependence on predefined layout templates limits user-friendliness. Optimization-based methods such as \cite{tang2024diffuscene, chen2024comboverse, rahamim2024lay, han2024reparo, zhou2024zero, tang2025zeroscene} progressively optimize object spatial positions to align with image priors, yet their generation stability remains problematic, often exhibiting deviations in object placement and orientation. Epstein et al. \cite{epstein2024disentangled} independently optimize object representations and learn spatial affine transformation parameters to construct diverse layouts. While Gen3DSR  \cite{Ardelean2025Gen3DSR} assembles reconstructed individual 3D objects into scenes using monocular depth guidance, its generated geometry and texture quality exhibit limitations. Recent works, MIDI \cite{huang2024midi}, introduces a multi-instance attention mechanism that directly captures spatial relationships between objects during the diffusion process and CAST \cite{yao2025castcomponentaligned3dscene} achieves component-wise generative alignment between canonical and scene spaces by computing model transformations, enabling efficient and accurate 3D scene generation from single images. 

Layout generation constitutes a critical research direction in compositional scene synthesis. Several methods \cite{feng2023layoutgpt, ocal2024sceneteller, zhang2024towards, zhou2024gala3d, li2024dreamscene, feng2025text, zhou2025layoutdreamer, sun2024layoutvlm} leverage the powerful semantic understanding of large language models to extract scene elements and their interrelations directly from user prompts, generating plausible coarse layouts. However, these approaches struggle to capture precise geometric and physical constraints. Scene graph-based methods \cite{dhamo2021graph, zhai2023commonscenes, zhai2024echoscene, yang2025mmgdreamer, hu2024scenecraft, gao2024graphdreamer, li2025phip} provide intuitive structured relationships but face limitations in accurate spatial relationship modeling. By leveraging the prior knowledge provided by images, our method circumvents the inaccuracies introduced by text or scene graphs, thereby enabling the construction of scenes that maintain high consistency with reference images.

\section{METHOD}

In this section, we present a detailed description of the pipeline for converting a single image into a structured 3D scene, which is decomposed into three collaborative subtasks: instance segmentation and generation, point cloud extraction, and layout optimization. The overview of the proposed framework is shown in Fig. \ref{Overview}.

\subsection{Instance Segmentation and Generation}\label{subtask1}

Corresponding to a given image, our method begins by generating individual 3D assets with high-quality geometric and texture details as well as multi-view consistency through a segmentation-reconstruction pipeline. Specifically, we first perform foreground object detection on the input single image $I$, identifying candidate targets by establishing associations between image features and predefined semantic labels $\mathcal{S}$. This process generates bounding boxes $b$, category labels $l$, and confidence scores $\alpha$ for $N$ instances in the scene, which can be mathematically expressed as:
\begin{equation}
    \text{Object Detection}(I, \mathcal{S}) \rightarrow \{b_i, l_i, \alpha_i\}_{i=1}^N, \ \ \alpha_i > \theta
    \label{eq1}
\end{equation}
Here, $i$ denotes the instance index, and $\alpha_i$ represents the matching degree between the instance and its assigned category $l_i$. To ensure subsequent processing quality, we retain only detection results with confidence scores exceeding a predefined threshold $\theta$, as low-confidence instances typically suffer from severe occlusion or category misclassification. Subsequently, for these high-confidence candidate regions bounded by $b$, we employ a refined segmentation module for pixel-level optimization. Through a bounding box-guided segmentation mechanism, we generate precise segmented images $p$ and corresponding binary mask matrices $m$ for the identified $N$ instances. This step is formulated as:
\begin{equation}
    \text{Instance Segmentation}(I, b_i, l_i, \alpha_i) \rightarrow \{p_i, m_i\}_{i=1}^N
    \label{eq2}
\end{equation}

\begin{figure}
    \centering
    \vspace{-0.4cm}
    \includegraphics[width=\columnwidth]{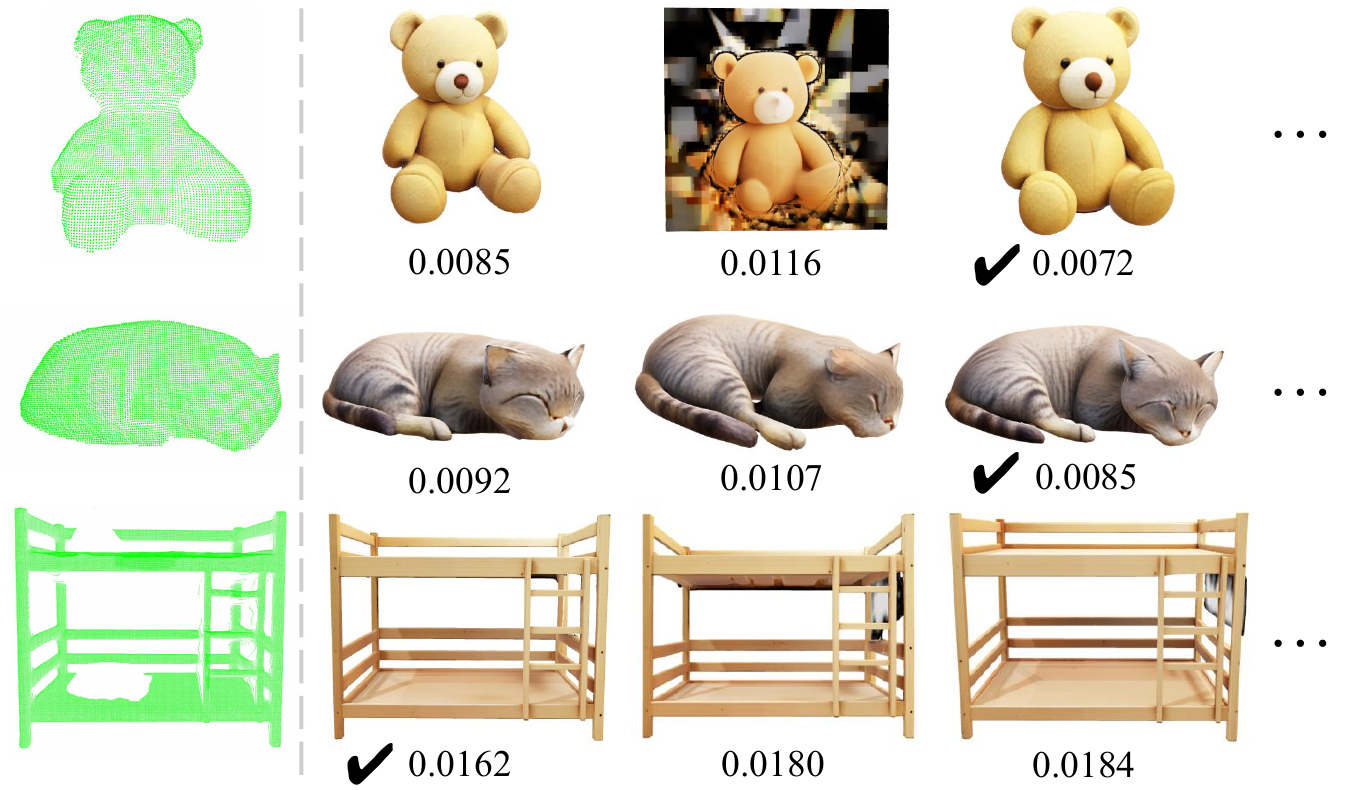}
    %\vspace{-0.7cm}
    \caption{\textbf{Model Selection Strategy.} We sample the multiple generated models into point cloud representations, and evaluate their quality by calculating the Chamfer Distance between them and the extracted instance point clouds (lower values correspond to superior models). The optimal model is subsequently selected for scene assembly in subtask 3.}
    \vspace{-0.4cm}
    \label{Model Selection Strategy.}
\end{figure}

Benefiting from the customizable semantic label set $\mathcal{S}$, the above pipeline allows flexible specification of target object categories according to specific requirements. This enables efficient filtering and acquisition of desired foreground instance images with high-quality masks, laying a robust foundation for subsequent generation tasks.

As shown in Fig. \ref{Overview}, original instance images often contain holes caused by mutual occlusion between objects, and these defects directly compromise the quality of 3D generation. To address this, we involve an inpainting phase before the image to 3D object generation step. During the inpainting phase, we leverage the superior semantic understanding capabilities of Vision-Language Models (VLMs) to visually localize defective regions in images through text prompts, generating inpainted images that effectively preserve the structural integrity of objects. In this work, we employ the VLM GPT-4o \cite {hurst2024gpt}, guiding the model via prompts to function as a professional inpainting system, and the outputs include the description of the main subject of the image, the inference of damaged regions, and the generation of inpainting result.

After completing the foreground instance segmentation images collection $\{p_i\}_{i=1}^N$, the powerful generative capability of Trellis \cite{xiang2024structured} assists us in modeling 3D assets from the inpainted image. Initially, we extract visual features with rich semantic information and partial 3D awareness from a single input image. These features are subsequently injected through cross-attention layers to progressively denoise and generate a low-resolution feature grid, which is further converted into active voxels representing the coarse contour and structure of the 3D object. Building upon this foundation, the model generates corresponding dense local latent vectors. Together, these components form a unified structured latent representation that comprehensively captures both geometric and appearance characteristics of the object. This flexible structured latent representation can then be efficiently decoded into diverse 3D formats. Through this pipeline, we map the single-view inpainted image to a collection of $K$ candidate models $\{\mathcal{M}_i^k\}_{k=1}^K$, each comprising both mesh and point cloud representations. 

\begin{figure}
    \centering
    \vspace{-0.4cm}
    \includegraphics[width=\columnwidth]{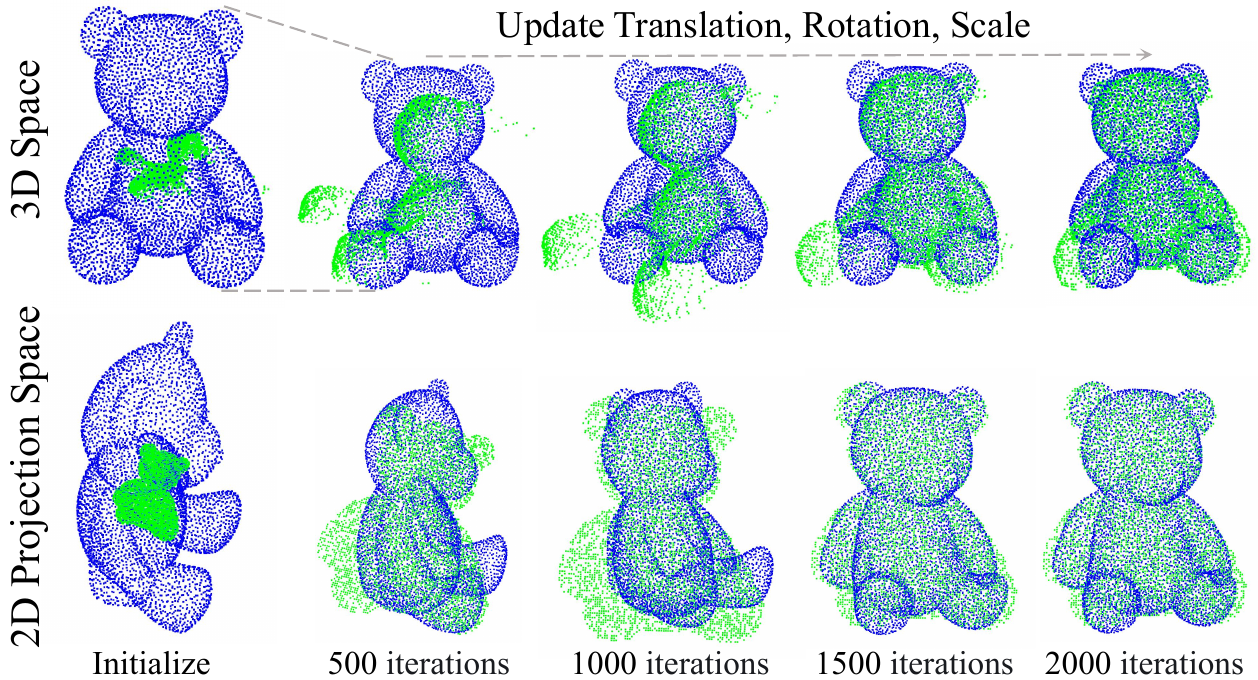}
    %\vspace{-0.7cm}
    \caption{\textbf{Layout optimization process.} Taking the parameter optimization of a bear toy as an example, the \textcolor{blue}{blue} point cloud represents the point cloud of the optimal model, i.e., the object to be optimized, and the \textcolor{green}{green} point cloud is the instance point cloud extracted in Sec. \ref{subtask2}, i.e., the target. We visualize the optimization process in both 3D space and 2D projection space, and obtain better layout parameters by integrating the information from the dual spaces.}
    \vspace{-0.4cm}
    \label{Layout optimization process.}
\end{figure}

\subsection{Point cloud Extraction}\label{subtask2}

The second stage of our generation framework involves the extraction of scene point clouds, which provides critical reference for subsequent 3D model registration and layout optimization. To recover the 3D point cloud representation of a scene from uncalibrated images, the key lies in accurately estimating the depth information of the images and the corresponding camera parameters, so as to establish the mapping from 2D pixel coordinates to 3D spatial coordinates.

Specifically, we construct a pseudo-stereo input pair using the original image $I$ and its copy $I_c$, which are processed through a shared vision encoder to extract feature representations. These features are subsequently fed into two independent decoders within the pre-trained deep learning module, DUSt3R \cite{wang2024dust3r}. The decoders continuously exchange information via cross-attention layers, enabling mutual reasoning between views, which is a critical mechanism for aligning their 3D representations. In the final stage, network regresses a 3D pointmap, defined as a 2D field of dense 3D points. From this pointmap, we directly extract a point cloud $\mathcal{PC}$ that contains the complete geometric structure of the scene in the camera coordinate. Notably, the $z$-axis coordinates of the pointmap inherently form a depth map $\mathcal{D}$. Under assumptions of principal point centering and square pixels, camera parameters $\mathcal{C}$ can be estimated by optimizing the reprojection error between 3D points in the pointmap and their corresponding 2D pixel positions. By integrating the generated pointmap with instance-level masks $m_i$ obtained from subtask 1, we perform spatial segmentation to produce independent point cloud representations $\mathcal{PC}_i$ for each identified instance. This processing pipeline can be formally expressed as:
\begin{equation}
    \{I,I_c\} \rightarrow \{ \mathcal{PC},\mathcal{D},\mathcal{C}\} \ \xrightarrow{with \ m_i} \ \{\mathcal{PC}_i\}_{i=1}^N
    \label{eq3}
\end{equation}

In order to overcome the instability of the generation results from the instance generation stage, i.e., Sec. \ref{subtask1}, we propose a model selection strategy based on the normalized Chamfer distance. As shown in Fig. \ref{Model Selection Strategy.}, coordinate normalization is applied to align the candidate point cloud $\mathcal{M}_i^k$ with the instance point cloud $\mathcal{PC}_i$. Then, the bidirectional Chamfer distance is computed as:
\begin{equation}
\begin{split}
CD(\mathcal{M}_i^k, \mathcal{PC}_i) &= \frac{1}{|\mathcal{M}_i^k|}\sum_{x\in\mathcal{M}_i^k}\min_{y\in\mathcal{PC}_i}\|x-y\|_2^2 \\
&+ \frac{1}{|\mathcal{PC}_i|}\sum_{y\in\mathcal{PC}_i}\min_{x\in\mathcal{M}_i^k}\|y-x\|_2^2
\end{split}
\label{eq4}
\end{equation}
where $|\cdot|$ denotes the cardinality of the point cloud. By minimizing this metric, we select the optimal model $\mathcal{M}_i = \arg\min_k CD(\mathcal{M}_i^k, \mathcal{PC}_i)$ as the final 3D representation. This approach effectively ensures geometric consistency between generated models and the original scene, establishing a solid foundation for final scene composition.

\begin{figure*}[!ht]
    \centering
    \vspace{-0.5cm}
    \includegraphics[width=\textwidth]{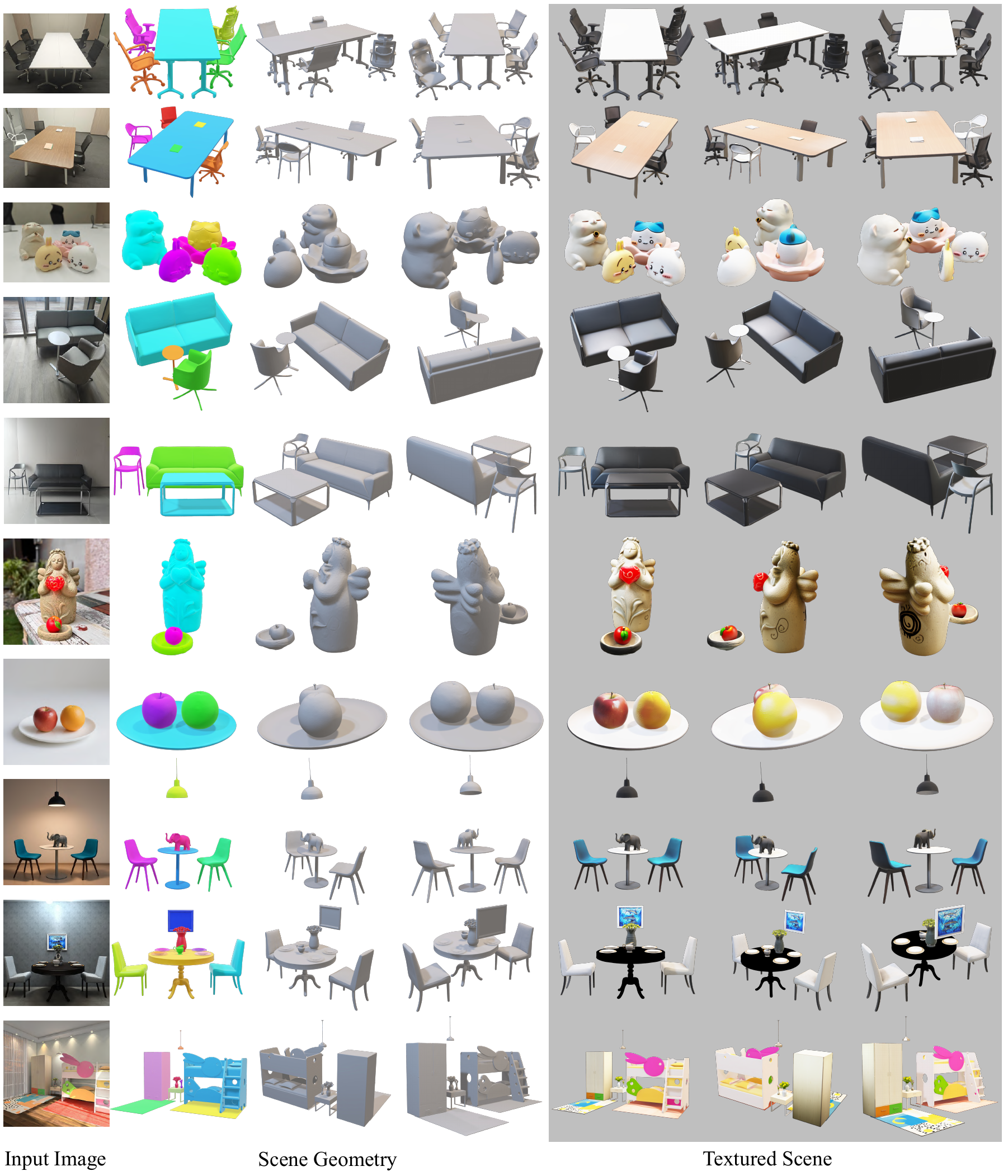}
    \vspace{-0.7cm}
    \caption{\textbf{Scene generation results of our method.} From left to right: input image, scene geometry (independent assets are annotated with distinct colors), and textured scene. The images in the first column from top to bottom are sourced from: real photographs (rows 1-5), VLM generated images (rows 6-7), 3D-FRONT \cite{fu20213d} (rows 8-10).}
    \vspace{-0.3cm}
    \label{presentation}
\end{figure*}

\subsection{Layout Optimization}\label{subtask3}

After obtaining the optimal 3D representation for each instance, precise spatial arrangement of objects must be performed according to the scene layout of the original image. To achieve this, we parameterize each instance as a learnable parameter set $\phi = \{T, R, S\}$ in 3D space, where the translation parameters $T = (T_x, T_y, T_z)$ represent object positions, the rotation parameters $R = (R_x, R_y, R_z)$ characterize object orientations, and the scaling parameter $S$ is an isotropic scaling factor.

The initialization strategy is crucial to the convergence speed and stability of optimization. To provide a favorable starting point for optimization, we adopt a geometric center-based initialization method. First, we calculate the centroid of the point cloud $\mathcal{PC}_i$ for each target instance, and initialize the translation parameter $T$ as the centroid coordinates, thereby roughly aligning each generative model $\mathcal{M}_i$ to the target region. The rotation parameter $R$ is initialized as an identity matrix, while the scaling parameter $S$ is initialized using the ratio of the bounding box diagonals for $\mathcal{M}_i$ and $\mathcal{PC}_i$. Subsequently, these parameters are optimized via gradient descent to maintain spatial consistency between 3D objects and the original image layout. Specifically, we adopt the evaluation criteria in Eq. \ref{eq4} as the core optimization objective, which minimizes the Chamfer Distance loss in 3D space between the generated point cloud $\mathcal{M}_i$ and the target instance point cloud $\mathcal{PC}_i$. It should be noted that $\mathcal{PC}_i$ only contains single-view information, namely the visible surface of the object facing the camera, and is geometrically incomplete. In contrast, the generated model $\mathcal{M}_i$ is a watertight mesh with a complete structure. Despite such structural asymmetry, we find that the visible points in $\mathcal{PC}_i$ are sufficient to provide dense correspondences for constraining the spatial parameters of the objects.

However, relying solely on 3D spatial constraints fails to achieve stable parameter convergence, as erroneous optimization could occur in the translation and rotation estimations of objects. Prior works \cite{chen2021unsupervised, zhou2024zero} introduced a 2D projection constraint mechanism that samples discrete points within the contour region of the original instance mask as supervision signals. By projecting the generated 3D point cloud onto multi-view 2D image planes to obtain projected point sets, this approach enforces geometric consistency across dimensions through a 2D Chamfer Distance loss to align the point sets. Inspired by this, we leverage camera parameters $\mathcal{C}$ to project both $\mathcal{M}_i$ and $\mathcal{PC}_i$ onto 2D planes and minimize their 2D Chamfer Distance loss. The final loss function is formulated as:
\begin{equation}
\begin{split}
    \mathcal{L} &= \lambda_1 \cdot \mathcal{L}_{CD}^{3D}(\mathcal{M}_i, \mathcal{PC}_i) \\
    &+ \lambda_2 \cdot \mathcal{L}_{CD}^{2D}(\text{Proj}_\mathcal{C}(\mathcal{M}_i), \text{Proj}_\mathcal{C}(\mathcal{PC}_i))
\end{split}
\label{eq5}
\end{equation}
where $\lambda_1$ and $\lambda_2$ are weighting coefficients balancing different supervision signals, $\text{Proj}_\mathcal{C}(\cdot)$ denotes the perspective projection operation based on the camera projection matrix. This 3D-2D joint optimization strategy effectively leverages the geometric information of visible surfaces, while compensating for the biases caused by the incompleteness of monocular depth point clouds through 2D contour constraints, thus ensuring dual consistency of the generated objects in both spatial layout and visual projection. Fig. \ref{Layout optimization process.} illustrates the optimization process of point cloud registration in both 3D and 2D spaces.

\section{EXPERIMENTS}

\subsection{Implementation Details.}

In the instance segmentation phase, i.e., Sec. \ref{subtask1}, our framework automatically performs object detection and pixel-level segmentation on the input image by leveraging Grounding DINO \cite{liu2024grounding} and SAM \cite{kirillov2023segment}. The confidence threshold in Eq. \ref{eq1} is set to $\theta=0.5$. We generate a diverse set of object candidates by setting the number of candidate models $K=5$ in the generation phase. In our layout optimization stage, the loss weights in Eq. \ref{eq5} are set to $\lambda_1=1$ and $\lambda_2 =5 \times 10^{-2}$. 

During the optimization phase, each instance undergoes 20 optimization epochs, with each epoch comprising 2,000 iterations. In the first 1,200 iterations, only the 3D Chamfer Distance loss is optimized to focus on spatial alignment of point clouds. For the subsequent 800 iterations, the 2D projection Chamfer Distance loss is introduced to enforce image-plane projection alignment constraints. The epoch with the minimum loss value is selected as the final optimized result. The Adam optimizer is adopted with a learning rate of 0.01 for translation, rotation, and scaling parameters. Through multiple scenario tests, we obtain the average computational time of each subtask, as summarized in Table \ref{subtasks time consuming}. All experiments are conducted on a single NVIDIA A100 GPU with 40GB of memory.

\begin{table}[!ht]
  \centering
  \caption{Average time consumption of each subtask.}
  \begin{tabular}{lccc}
    \toprule
    \textbf{} & \textbf{Subtask1}  & \textbf{Subtask2}  & \textbf{Subtask3} \\
    \midrule
    Time & 149.32s & 21.33s & 209.98s \\
    \bottomrule
  \end{tabular}
  \vspace{-0.2cm}
  \label{subtasks time consuming}
\end{table}

\subsection{Results}

Since most image-guided generation tasks treat the background as non-interactive, we validated our proposed method on a constructed image set containing multiple mutually occluded foreground objects. The results generated by our method are presented in Fig. \ref{presentation}. The data sources include 15 real-world photographs, 15 images generated by the VLM \cite{hurst2024gpt}, and 20 indoor synthetic scenes from the public benchmark 3D-FRONT \cite{fu20213df}. This dataset covers common indoor instance categories, primarily including tables, chairs, sofas, cabinets, lamps, and toys, with the objects in test images exhibit varying degrees of occlusion. To quantify the scene complexity, we analyze the spatial relationships between objects by leveraging the object detection results from subtask 1, and find that the average 2D bounding box intersection over union (IoU) among interactive instances reaches 16.09\%. Considering that bounding boxes are difficult to capture precise object shapes and that IoU is sensitive to object size, we regard an IoU range of 10\%–20\% as a moderate occlusion level, which ensures that the dataset can sufficiently challenge the model’s capability to recover occluded geometric shapes and optimize scene layouts.

For experimental validation, we adopted a hybrid evaluation protocol. Qualitative evaluation was conducted on the full dataset to demonstrate the generalization capability of our method across both real and generated domains. For quantitative comparison, we calculated the mean values of the metrics based on the 3D ground-truth scenes provided by 3D-FRONT. In addition, we carried out user studies and ablation experiments.

\noindent \textbf{Qualitative Comparison.} We conducted a comparative evaluation of scene generation results, selecting state-of-the-art methods as benchmarks, including MIDI \cite{huang2024midi}, Zhou et al. \cite{zhou2024zero}, Gen3DSR \cite{Ardelean2025Gen3DSR}, and CAST \cite{yao2025castcomponentaligned3dscene}. With the exception of CAST, all comparative methods were implemented using their publicly available source codes. Our evaluation focuses on two dimensions: object accuracy and scene layout. Fig. \ref{qualitative comparison2} demonstrates the results of each method under identical scene inputs. In terms of generation quality, MIDI only produces geometric meshes while lacking texture information, and it tends to suffer from morphological deviations when handling small objects. The method proposed by Zhou et al. exhibits obvious shape distortion and incomplete object reconstruction. Gen3DSR performs poorly in detail preservation. In contrast, both CAST and our method effectively maintain the structural integrity of individual objects. Regarding layout construction, MIDI yields erroneous depth estimation in some test scenarios, which leads to abnormal positions of individual objects. Neither Zhou et al.’s method nor Gen3DSR can effectively optimize object rotation parameters. CAST also fails to accurately recover the layout positions of all instances in certain cases. Our approach achieves precise modeling of spatial relationships between objects and ensures multi-view consistency across the entire scene.

\begin{table}[t]
\centering
\caption{\textbf{Quantitative comparison.} Since the assets generated by MIDI lack textural information, the CLIP-Score of color cannot be calculated and is denoted as ``-". A higher CLIP-Score indicates that the generated results have a greater correlation with the reference image, and the geometric and texture quality of the model is higher. A smaller Chamfer distance suggests that the spatial distance between the results and the reference scene is smaller, and the layout is more accurate. A higher F-Score represents greater reconstruction accuracy of the result.}
\vspace{-0.3cm}
\setlength{\tabcolsep}{1.5pt}
\resizebox{\linewidth}{!}{
\begin{tabular}{lcccccc} 
\toprule
\multirow{2}{*}{\textbf{Methods}} & \multicolumn{2}{c}{\textbf{CLIP-Score}} & \multicolumn{2}{c}{\textbf{Chamfer Distance}} & \multicolumn{2}{c}{\textbf{F-Score}} \\
\cmidrule(lr){2-3} \cmidrule(lr){4-5}  \cmidrule(lr){6-7}
& \textbf{Geometry$\uparrow$} & \textbf{Color$\uparrow$} & \textbf{3D Space$\downarrow$} & \textbf{2D Space$\downarrow$} & \textbf{3D Space$\uparrow$} & \textbf{2D Space$\uparrow$}\\

\midrule
MIDI \cite{huang2024midi} & 0.8171 & - & 0.0143 & 5.8452 & 71.39 & 40.58 \\
Zhou et al. \cite{zhou2024zero} & 0.7606 & 0.7914 & 0.0224 & 6.9914 & 64.71 & 35.35\\
Gen3DSR \cite{Ardelean2025Gen3DSR} & 0.7682 & 0.8271 & 0.0189 & 6.1536 & 65.15 & 35.55\\
Ours & \textbf{0.8389} & \textbf{0.8990} & \textbf{0.0127} & \textbf{4.9264} & \textbf{76.60} & \textbf{44.12}\\

\bottomrule
\end{tabular}}
\vspace{-0.4cm}
\label{quantitative comparison}
\end{table}

\noindent \textbf{Quantitative Comparison.} We adopt CLIP-Score \cite{taited2023CLIPScore} as an evaluation metric to measure the correlation between rendered images and reference images. Specifically, we calculate CLIP scores for both white models and textured meshes respectively, thereby assessing the geometric and texture generation quality of different methods. We employ Chamfer Distance between point clouds as another crucial metric to quantify the spatial discrepancy between generated scenes and reference scenes. Additionally, the F-Score is incorporated into our evaluation framework. It comprehensively assesses the reconstruction accuracy and matching degree of object geometry, with thresholds set at 0.01 in 3D space and 1.00 in the 2D projection space. We continue to use the methods from the qualitative comparison, including MIDI, Zhou et al., Gen3DSR and our approach to conduct a quantitative analysis on identical input scenes. Since the source code and scene models of CAST are unavailable, this method is not included in the present quantitative comparison. As demonstrated in Table \ref{quantitative comparison}, our method exhibits superior performance compared to other approaches.

\begin{figure*}[t]
    \centering
    \vspace{-0.8cm}
    \includegraphics[width=\textwidth]{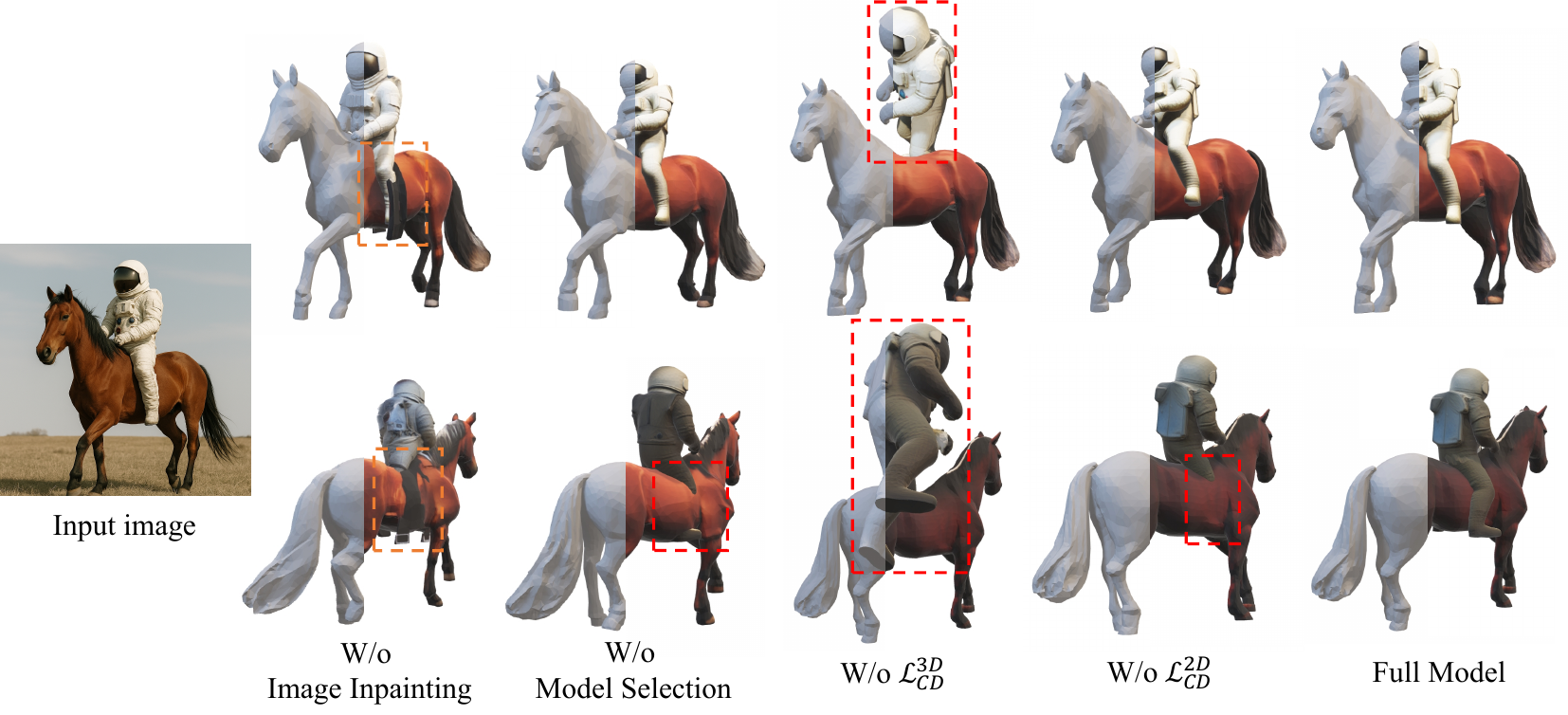}
    \vspace{-0.7cm}
    \caption{\textbf{Qualitative analysis of the ablation study on the removal of specific components.}
    }
    \vspace{-0.3cm}
    \label{Ablation Study Figure}
\end{figure*}

\begin{figure}[t]
    \centering
    \includegraphics[width=\columnwidth]{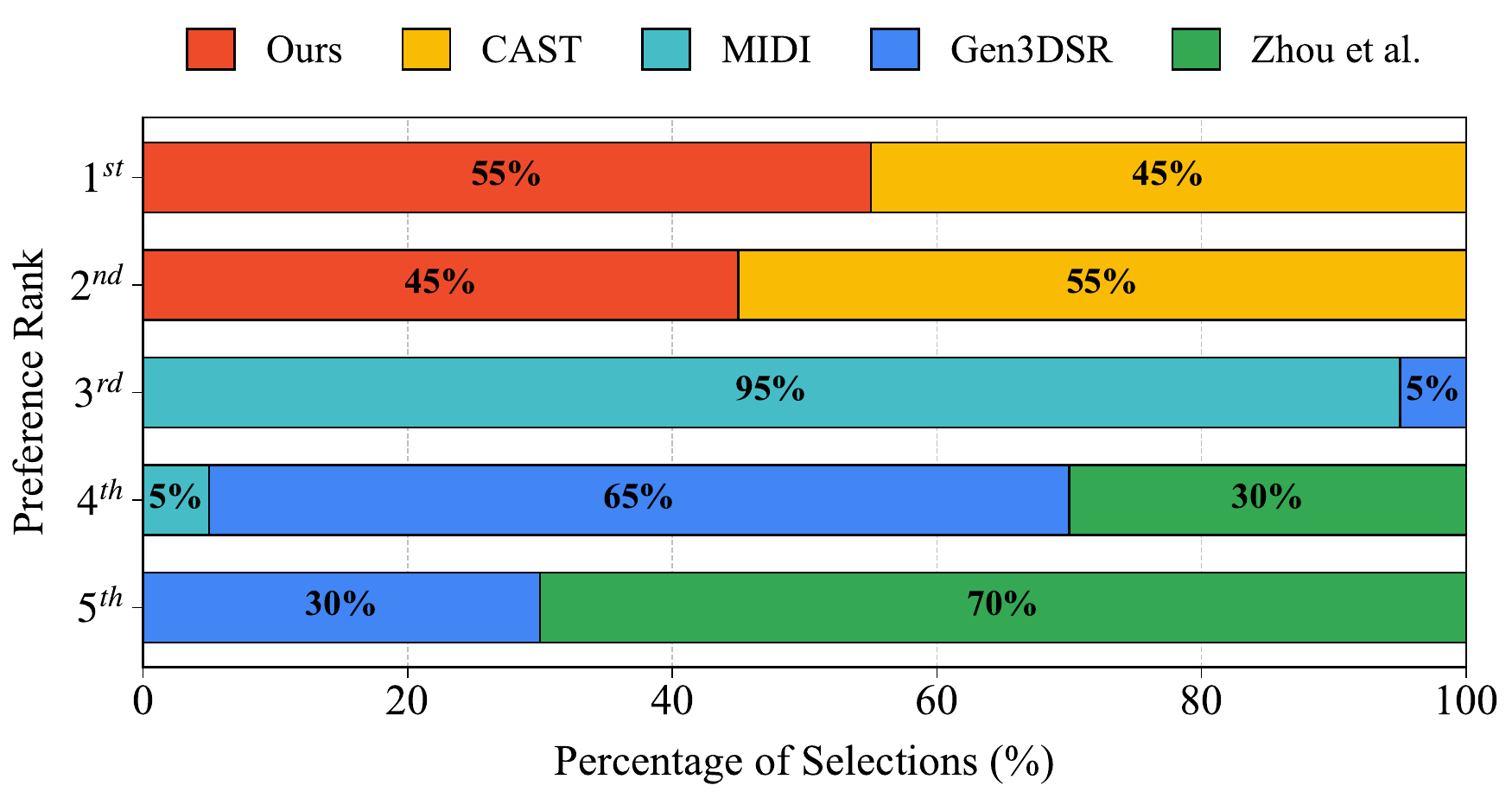}
    \vspace{-0.7cm}
    \caption{\textbf{User Study.} Percentages represent the proportion of cases in which each respective method achieves the corresponding ranking.}
    \vspace{-0.5cm}
    \label{User Study}
\end{figure}

\noindent \textbf{User Study.} We conducted a user study to perform a comparative analysis between our method and existing generative approaches. Participants were asked to evaluate multiple sets of scene rendering results, where each group contained a reference image and corresponding images generated by different methods for the same scene. Users were required to select the generated result that most closely resembled the reference image based on two criteria: object reconstruction accuracy and scene layout fidelity. To mitigate assessment bias, all options were presented in randomized order. We collected 400 responses from 40 human volunteers. As demonstrated in Fig. \ref{User Study}, our method outperforms most existing approaches in terms of human preference, attaining the highest votes in 55\% of the samples and slightly surpassing CAST.

\begin{table}[t]
\centering
\caption{\textbf{Quantitative analysis of the ablation study on the removal of specific components.}}
\vspace{-0.3cm}
\setlength{\tabcolsep}{1.5pt}
\resizebox{\linewidth}{!}{
\begin{tabular}{lcccccc} 
\toprule
\multirow{2}{*}{\textbf{Ablation}} & \multicolumn{2}{c}{\textbf{CLIP-Score}} & \multicolumn{2}{c}{\textbf{Chamfer Distance}} & \multicolumn{2}{c}{\textbf{F-Score}} \\
\cmidrule(lr){2-3} \cmidrule(lr){4-5}  \cmidrule(lr){6-7}
& \textbf{Geometry$\uparrow$} & \textbf{Color$\uparrow$} & \textbf{3D Space$\downarrow$} & \textbf{2D Space$\downarrow$} & \textbf{3D Space$\uparrow$} & \textbf{2D Space$\uparrow$}\\
\midrule
W/o Inp. & 0.7965 & 0.8568 & 0.0219 & 6.1589 & 73.28 & 38.24\\*[\dimexpr\arrayrulewidth+2pt]
W/o Sel. & 0.7929 & 0.8467 & 0.0236 & 5.9385 & 72.52 & 38.77\\*[\dimexpr\arrayrulewidth+2pt]
W/o $\mathcal{L}_{CD}^{3D}$ & 0.7287 & 0.7729 & - & 6.2748 & - & 39.71 \\*[\dimexpr\arrayrulewidth+2pt]
W/o $\mathcal{L}_{CD}^{2D}$ & 0.8058 & 0.8495 & 0.0194 & - & 73.13 & -\\*[\dimexpr\arrayrulewidth+2pt]
Full Model & \textbf{0.8389} & \textbf{0.8990} & \textbf{0.0127} & \textbf{4.9264} & \textbf{76.60} & \textbf{44.12}\\
 
\bottomrule
\end{tabular}}
\vspace{-0.3cm}
\label{Ablation Study table}
\end{table}

\subsection{Ablation Study}

To comprehensively evaluate the effectiveness of the proposed framework, we designed two categories of ablation studies. The first category verifies the necessity of specific components in the system by removing them, while the second category demonstrates the superiority of the current design choices by replacing key modules with alternative counterparts of the same type.

\begin{figure*}[!ht]
    \centering
    \vspace{-0.8cm}
    \includegraphics[width=\textwidth]{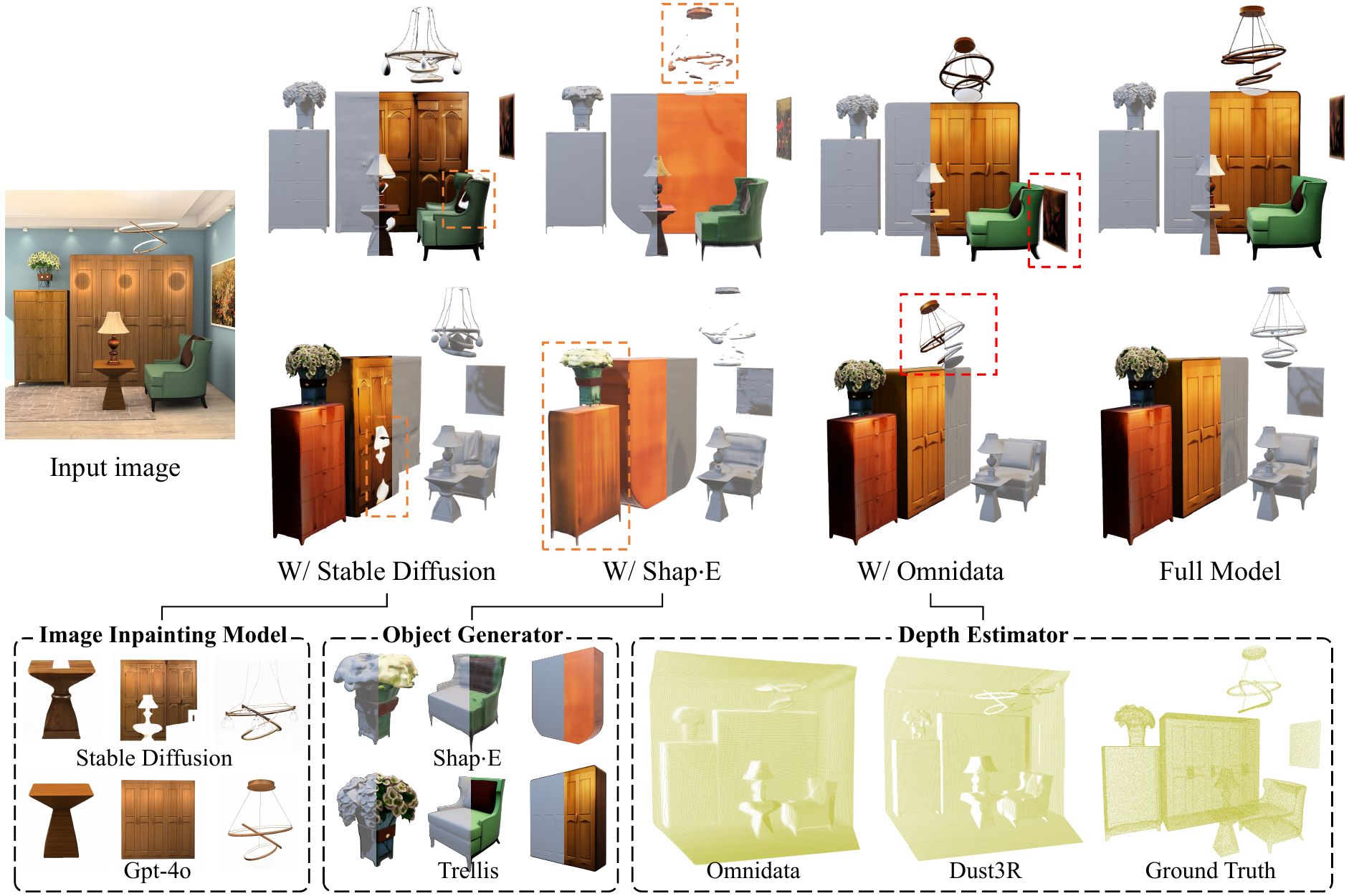}
    \vspace{-0.7cm}
    \caption{\textbf{Qualitative analysis of the ablation study on the replacement of key modules.}
    }
    \vspace{-0.1cm}
    \label{Ablation Study Figure 2}
\end{figure*}

\begin{table*}[!ht]
\centering
\caption{\textbf{Quantitative analysis of the ablation study on the replacement of key modules.}}
\vspace{-0.3cm}
\setlength{\tabcolsep}{2pt}
\resizebox{\linewidth}{!}{
\begin{tabular}{lcccccccccccc} 
\toprule
\multirow{2}{*}{\textbf{Ablation}} & \multicolumn{2}{c}{\textbf{Inpainting Model}} & \multicolumn{2}{c}{\textbf{Object Generator}} & \multicolumn{2}{c}{\textbf{Depth Estimator}} & \multicolumn{2}{c}{\textbf{CLIP-Score}} & \multicolumn{2}{c}{\textbf{Chamfer Distance}} & \multicolumn{2}{c}{\textbf{F-Score}} \\
\cmidrule(lr){2-3} \cmidrule(lr){4-5}  \cmidrule(lr){6-7} \cmidrule(lr){8-9} \cmidrule(lr){10-11} \cmidrule(lr){12-13}
& \textbf{Stable Diffusion} & \textbf{Gpt-4o} & \textbf{Shap$\cdot$E} & \textbf{Trellis} & \textbf{Omnidata} & \textbf{Dust3R} & \textbf{Geometry$\uparrow$} & \textbf{Color$\uparrow$} & \textbf{3D Space$\downarrow$} & \textbf{2D Space$\downarrow$} & \textbf{3D Space$\uparrow$} & \textbf{2D Space$\uparrow$}\\
\midrule
W/ Stable Diffusion & \usym{2714} & \boldmath $-$ & \boldmath $-$ & \usym{2714} & \boldmath $-$ & \usym{2714} & 0.8139 & 0.8421 & 0.0162 & 5.1537 & 74.19 & 40.08\\*[\dimexpr\arrayrulewidth+2pt]
W/ Shap$\cdot$E & \boldmath $-$ & \usym{2714} & \usym{2714} & \boldmath $-$ & \boldmath $-$ & \usym{2714} & 0.7826 & 0.8183 & 0.0195 & 6.3021 & 68.82 & 37.26\\*[\dimexpr\arrayrulewidth+2pt]
W/ Omnidata & \boldmath $-$ & \usym{2714} & \boldmath $-$ & \usym{2714} & \usym{2714} & \boldmath $-$ & 0.8053 & 0.8506 & 0.0218 & 6.9753 & 72.57 & 39.90 \\*[\dimexpr\arrayrulewidth+2pt]
Full Model & \boldmath $-$ & \usym{2714} & \boldmath $-$ & \usym{2714} & \boldmath $-$ & \usym{2714} & \textbf{0.8389} & \textbf{0.8990} & \textbf{0.0127} & \textbf{4.9264} & \textbf{76.60} & \textbf{44.12}\\
 
\bottomrule
\end{tabular}}
\vspace{-0.5cm}
\label{Ablation Study table 2}
\end{table*}

% \begin{table*}[!ht]
% \centering
% \caption{\textbf{Quantitative analysis of the ablation study on the replacement of key modules.}}
% \setlength{\tabcolsep}{4pt}
% \resizebox{\linewidth}{!}{
% \begin{tabular}{cccccccccccc} 
% \toprule
% \multicolumn{3}{c}{\textbf{Usage of Key Modules}} & \multicolumn{2}{c}{\textbf{CLIP-Score}} & \multicolumn{2}{c}{\textbf{Chamfer Distance}} & \multicolumn{2}{c}{\textbf{F-Score}} \\
% \cmidrule(lr){1-3} \cmidrule(lr){4-5} \cmidrule(lr){6-7} \cmidrule(lr){8-9}
% \textbf{Image Inpainting Model} & \textbf{Object Generator} & \textbf{Depth Estimator} & \textbf{Geometry$\uparrow$} & \textbf{Color$\uparrow$} & \textbf{3D Space $\downarrow$} & \textbf{2D Space $\downarrow$} & \textbf{3D Space $\uparrow$} & \textbf{2D Space $\uparrow$}\\
% \midrule
% Stable Diffusion & Trellis & Dust3R & 0.8139 & 0.8421 & 0.0162 & 5.1537 & 74.19 & 40.08\\*[\dimexpr\arrayrulewidth+2pt]
% Gpt-4o & Shap$\cdot$E & Dust3R & 0.7826 & 0.8183 & 0.0195 & 6.3021 & 68.82 & 37.26\\*[\dimexpr\arrayrulewidth+2pt]
% Gpt-4o & Trellis & Omnidata & 0.8053 & 0.8506 & 0.0218 & 6.9753 & 72.57 & 39.90 \\*[\dimexpr\arrayrulewidth+2pt]
% \textbf{Gpt-4o} & \textbf{Trellis} & \textbf{Dust3R} & \textbf{0.8389} & \textbf{0.8990} & \textbf{0.0127} & \textbf{4.9264} & \textbf{76.60} & \textbf{44.12}\\
 
% \bottomrule
% \end{tabular}}
% \vspace{-0.5cm}
% \label{Ablation Study table 2}
% \end{table*}

\begin{table}[t]
\centering
\caption{\textbf{Quantitative Comparison.} ``Ours$^*$" denotes a variant of our method where key modules are replaced with configurations consistent with Zhou et al. \cite{zhou2024zero}.}
\vspace{-0.3cm}
\setlength{\tabcolsep}{1.5pt}
\resizebox{\linewidth}{!}{
\begin{tabular}{lcccccc} 
\toprule
\multirow{2}{*}{\textbf{Method}} & \multicolumn{2}{c}{\textbf{CLIP-Score}} & \multicolumn{2}{c}{\textbf{Chamfer Distance}} & \multicolumn{2}{c}{\textbf{F-Score}} \\
\cmidrule(lr){2-3} \cmidrule(lr){4-5}  \cmidrule(lr){6-7}
& \textbf{Geometry$\uparrow$} & \textbf{Color$\uparrow$} & \textbf{3D Space$\downarrow$} & \textbf{2D Space$\downarrow$} & \textbf{3D Space$\uparrow$} & \textbf{2D Space$\uparrow$}\\
\midrule
Zhou et al. \cite{zhou2024zero} & 0.7606 & 0.7914 & 0.0224 & 6.9914 & 64.71 & 35.35\\
Ours$^*$ & \textbf{0.7662} & \textbf{0.8025} & \textbf{0.0201} & \textbf{6.9907} & \textbf{65.16} & \textbf{35.97}\\
\bottomrule
\end{tabular}}
\vspace{-0.4cm}
\label{compare}
\end{table}

We conducted ablation studies on the image inpainting and model selection modules within the framework, as well as different combinations of the loss function in Eq. \ref{eq5}. Fig. \ref{Ablation Study Figure} presents a qualitative comparison using a sample case, while Table \ref{Ablation Study table} provides a quantitative analysis over more samples. The experimental results demonstrate that directly using un-inpainted segmented images for generation leads to 3D assets with redundant geometries and erroneous poses, which subsequently adversely affect the layout optimization. Furthermore, omitting the model selection strategy results in randomly sampled models whose compatibility with the instance point cloud cannot be guaranteed, thereby introducing interference into the final outcome. When solely employing $\mathcal{L}_{CD}^{2D}$ as the loss function, the absence of 3D spatial depth constraints leads to severe positional misalignment of objects during optimization, resulting in significant deviations from reference data in metrics such as CLIP-Score and Chamfer Distance. When exclusively using $\mathcal{L}_{CD}^{3D}$, although the depth prior facilitates the capture of inter-object spatial relationships, the optimization process for object positions and rotation parameters still suffers from unstable convergence, causing model degradation to some extent.

To isolate the impact of each component, we performed replacement comparisons on the foundational models within our framework. Concretely, we replaced GPT-4o in the image inpainting stage with a diffusion-based inpainting model (i.e., Stable Diffusion \cite{rombach2022high}), substituted the single object generator Trellis with Shap$\cdot$E \cite{jun2023shap}, and swapped the monocular depth model DUSt3R for Omnidata \cite{eftekhar2021omnidata}, respectively. Fig. \ref{Ablation Study Figure 2} and Table \ref{Ablation Study table 2} present the qualitative and quantitative comparison results of the relevant scenarios. Experiments demonstrate that in terms of image inpainting, Stable Diffusion struggles to accurately infer the occluded structures of objects under the same prompts, thereby misleading subsequent generation tasks. Regarding 3D asset quality, Shap$\cdot$E exhibits weak geometry generation capability and only supports vertex colors while lacking fine-grained textures. In the aspect of depth estimation, compared with the pointmaps generated by DUSt3R, the 3D point clouds projected from Omnidata depth maps suffer from edge blurring and excessive noise. This results in sparse valid matching points during the layout optimization process, ultimately leading to object misalignment. Moreover, to further validate our framework, we designed a set of controlled experiments by replacing the aforementioned modules with configurations consistent with those of Zhou et al. \cite{zhou2024zero}. The results in Table \ref{compare} demonstrate that our method outperforms that of Zhou et al. on all metrics even when built upon the same base model. This superiority is mainly attributed to the 3D-2D joint optimization in layout refinement, which enables more effective utilization of prior information.

The ablation experiments verify that the full model, by integrating advanced foundational vision models, the image inpainting and model selection strategy, generates optimal 3D assets with consistent geometry and texture. By jointly optimizing both 2D projection constraints and 3D constraints, it effectively overcomes the limitations of a single loss function and achieves superior comprehensive performance across all evaluation metrics.
\section{DISCUSSION}

\begin{figure}
    \centering
    %\vspace{-0.3cm}
    \includegraphics[width=\columnwidth]{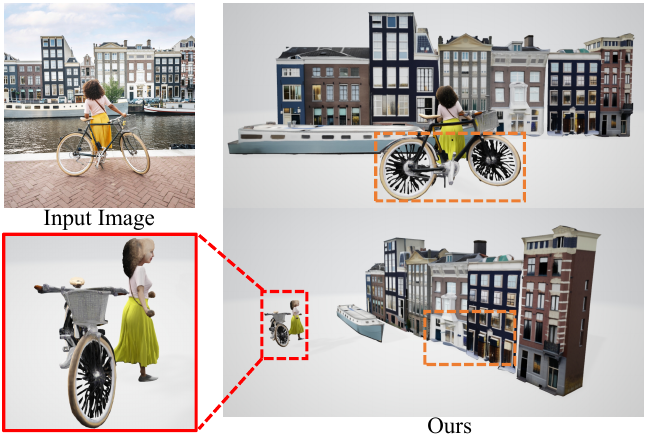}
    \vspace{-0.7cm}
    \caption{\textbf{Failure case.}}
    \vspace{-0.4cm}
    \label{Failure case.}
\end{figure}

This paper centers on the generation of room-level or tabletop scenes and exhibits favorable robustness to moderate occlusion. However, when the 2D bounding boxes IoU of objects exceeds 25\% (i.e., severe occlusion), the difficulty of instance image inpainting increases remarkably, which in turn impairs the subsequent generation and assembly of 3D assets. In terms of computational efficiency, our pipeline theoretically imposes no upper limit on the number of objects within a scene, yet the time overhead of layout optimization scales linearly with the number of objects. Moreover, when handling large-scale outdoor and in-the-wild scenes, severe occlusions and complex geometries tend to give rise to erroneous inference and generation of object shapes, which further impairs the layout optimization process and consequently leads to the degradation in performance of current methods, as illustrated in Fig. \ref{Failure case.}. We intend to extend our experiments to scenarios containing increased object density and interactions, thereby progressively expanding the scope and applicability of our research. Since most application scenarios for image-guided scene generation, such as XR-entertainment, embodied intelligence and autonomous navigation, emphasize the interaction with foreground objects, while the background is considered as non-interactive, our current pipeline treats the image background as an inactive object at infinity and does not participate in the 3D generation phase. For those scene construction tasks with complex backgrounds, for example, the game scene or urban scene modeling, our approach may give out unreliable results as there's no occlusion check between foreground objects and background in the layout optimization step. Thus, our immediate plan is to overcome the issue of excessive noise caused by the absence of decoupling between image background and foreground in the point cloud extraction process in order to estimate the correct background depth for layout optimization.

%In the point cloud extraction process, the absence of decoupling between image background and foreground, combined with inherent errors in camera parameters and depth map estimation, may lead to excessive noise artifacts or erroneous structures in the point clouds. Such distortions could adversely affect parameter convergence during layout optimization, potentially resulting in incorrect object placement. A potential solution involves fine-tuning diffusion models to generate multi-view images, thereby mitigating depth estimation errors through multi-view matching.

\begin{figure*}[ht]
    \centering
    \vspace{-0.9cm}
    \includegraphics[width=\textwidth]{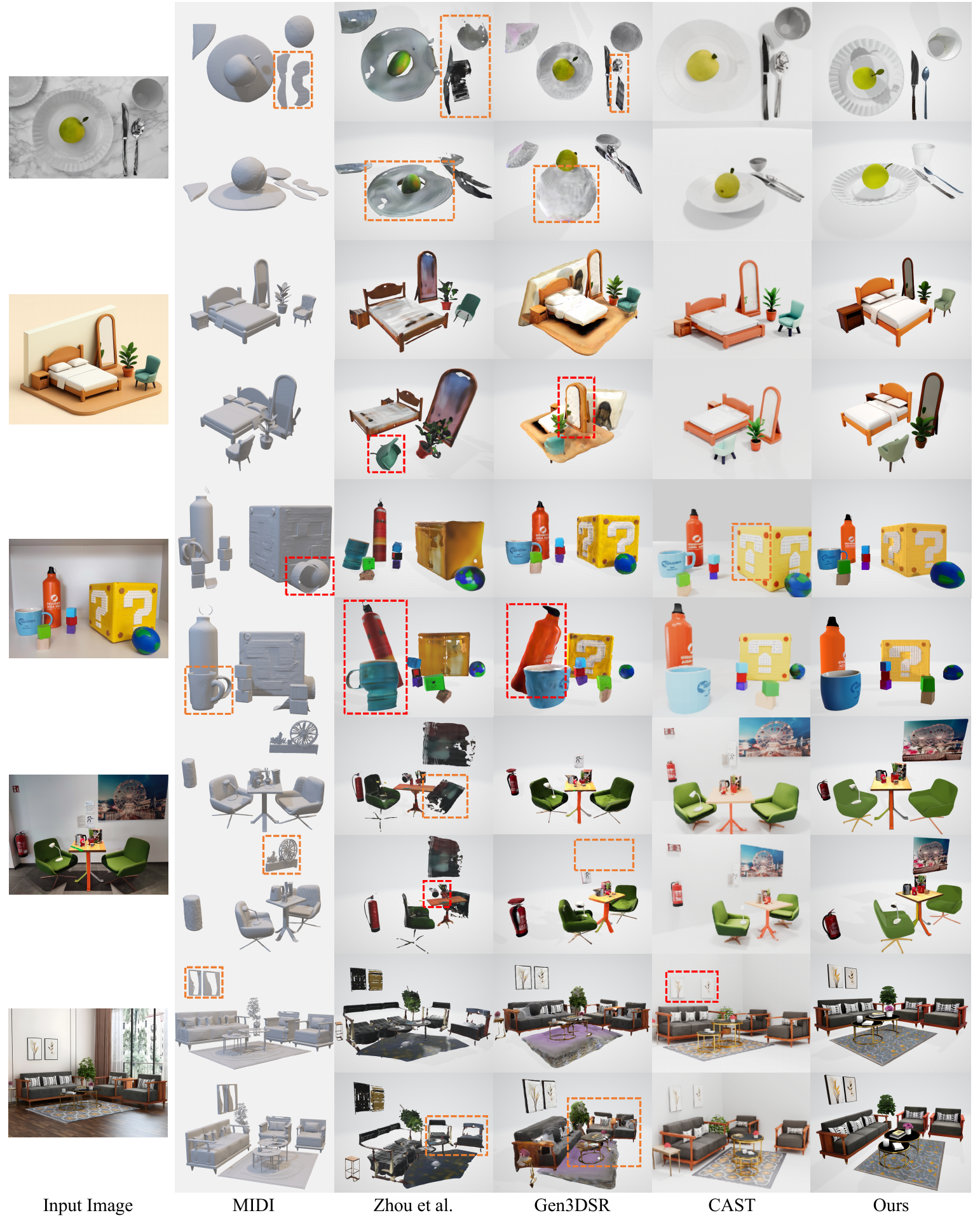}
    \vspace{-0.3cm}
    \caption{\textbf{Qualitative comparisons of single-image scene generation methods.} From left to right: input image, MIDI \cite{huang2024midi}, Zhou et al. \cite{zhou2024zero}, Gen3DSR \cite{Ardelean2025Gen3DSR}, CAST \cite{yao2025castcomponentaligned3dscene} (results are quoted from the original literature), and our approach. The images in the first column from top to bottom are sourced from: VLM generated images (rows 1-2), real photographs (rows 3-4),  and 3D-FRONT \cite{fu20213d} (row 5). \textcolor{orange}{Orange} boxes highlight geometric inaccuracies and detail loss in object generation quality, and \textcolor{red}{red} boxes indicate inconsistencies in scene layout compared to the reference image.}
    \vspace{-0.3cm}
    \label{qualitative comparison2}
\end{figure*}

Our work mainly focuses on ensuring the completeness of instance generation and the spatial accuracy of the layout, while paying less attention to textural refinement of generated models. From experimental results, we observed there are over-exposure or under-exposure issues in some of the results, e.g., the 7th scene in Fig. \ref{presentation} and the 2nd scene in Fig. \ref{qualitative comparison2}. Therefore, our future work will address texture mapping optimization and material property refinement for scene objects, while incorporating illumination conditions to enhance rendering quality.

\section{CONCLUSION}

In this paper, we present a single image-guided model generation and layout optimization framework to generate the 3D scene with precise textured meshes and spatial details. Our framework can be considered as a decomposition-composition approach: a 3D instance generation method via image decomposition and inpainting is proposed for guiding the foreground objects generation, which ensures the geometric and appearance accuracy even if the objects in the guidance image are occluded. An instance point cloud extraction phase collaborates with Chamfer distance loss minimization, which not only overcomes the instability of 3D instance generation, but also ensures the spatial layout of generated scene aligns highly with the reference image. Experimental results demonstrate that our framework achieves finer geometric modeling and texture generation at the object level compared to prior image-to-3D methods. Moreover, when handling complex object interactions and occlusions, our method can effectively maintain spatial rationality and multi-view consistency, showing that it has the potential to be applied to larger and more complex scenarios.

{
    \small
    \bibliographystyle{ieeenat_fullname}
    \bibliography{main}
}

% WARNING: do not forget to delete the supplementary pages from your submission 
% \setcounter{section}{0}  % 重置section编号
% \input{sec/X_suppl}

\end{document}